\documentclass[11pt, preprint]{aastex63}
\usepackage{bm}

\newcommand{\ii}{\mathrm{in}}
\newcommand{\oo}{\mathrm{out}}
\newcommand{\pp}{\mathrm{pl}}
\newcommand{\BBH}{{\rm\scriptscriptstyle BBH}}

\received{2019 December 4}
\revised{2020 January 10}
\accepted{2020 January 18}

\shorttitle{A strategy to search for an inner binary black hole}
\shortauthors{Hayashi, Wang \& Suto}
\graphicspath{{./}{figures/}}
\begin{document}

\title{A strategy to search for an inner binary black hole 
  from the motion of the tertiary star}

\correspondingauthor{Toshinori Hayashi}
\email{hayashi@utap.phys.s.u-tokyo.ac.jp}

\author[0000-0003-0288-6901]{Toshinori Hayashi}% (林　利憲)}
\affiliation{Department of Physics, The University of Tokyo,  
Tokyo 113-0033, Japan}

\author[0000-0002-5635-2449]{Shijie Wang}% (汪　士杰)}
\affiliation{Department of Physics, The University of Tokyo,  
Tokyo 113-0033, Japan}

\author[0000-0002-4858-7598]{Yasushi Suto}% (須藤 靖)}
\affiliation{Department of Physics, The University of Tokyo,  
Tokyo 113-0033, Japan}
\affiliation{Research Center for the Early Universe, School of Science,
The University of Tokyo, Tokyo 113-0033, Japan}

%%%%%%%%%%%%%%%%%%%%%%%%%%%%%%%%%%%%%%%%%%%%%%%%%%%%%%%%%%%%%%%%%
\begin{abstract}
There are several on-going projects to search for stars orbiting
around an invisible companion. A fraction of such candidates may be a
triple, instead of a binary, consisting of an inner binary black hole
(BBH) and an outer orbiting star.  In this paper, we propose a
methodology to search for a signature of such an inner BBH, possibly a
progenitor of gravitational-wave sources discovered by {\it LIGO},
from the precise radial velocity (RV) follow-up of the outer star.  We
first describe a methodology using an existing approximate RV formula
for coplanar circular triples.  We apply this method and constrain the
parameters of a possible inner binary objects in 2M05215658+4359220,
which consists of a red giant and an unseen companion.  Next we
consider co-planar but non-circular triples. We compute numerically
the RV variation of a tertiary star orbiting around an inner BBH,
generate mock RV curves, and examine the feasibility of the BBH
detection for our fiducial models. We conclude that the short-cadence
RV monitoring of a star-BH binary provides an interesting and
realistic method to constrain and/or search for possible inner BBHs.
Indeed a recent discovery of a star--BH binary system LB-1 may imply
that there are a large number of such unknown objects in our Galaxy,
which are ideal targets for the methodology proposed here.
\end{abstract}
%%%%%%%%%%%%%%%%%%%%%%%%%%%%%%%%%%%%%%%%%%%%%%%%%%%%%%%%%%%%%%%%%


\section{Introduction \label{sec:intro}}

The first direct detection of a gravitational wave (GW) from a binary
black-hole (BBH) merger \citep{Abbott2016} has convincingly
established the presence of very compact BBHs in the universe.  While
the origin, evolution and distribution of such BBHs are not yet
understood, several interesting scenarios have been proposed so far.
The binary evolution scenario
\citep[e.g.][]{Belczynski2002,Belczynski2007,Belczynski2012,Belczynski2016,Dominik2012,Dominik2013,Kinugawa2014,Kinugawa2016,Spera2019}
assumes that a fraction of massive stars form binary systems,
experiences supernovae, and finally evolves into compact binaries
including BBHs.  The dynamical formation scenario, on the
other hand, considers that stars and BHs located in dense clusters experience significant gravitational interactions, and eventually form BBHs ejected 
from the cluster regions mainly through binary-single, in addition to binary-binary, scattering encounters \citep[e.g.][]{Zwart2000,OLeary2009,Rodriguez2016,Tagawa2016}.
The primordial BH scenario \citep[e.g.][]{Sasaki2016,Sasaki2018,Bird2016} proposes
that numerous primordial BHs are formed in the very early universe and
a tiny fraction of them experiences close encounter and forms tight
BBHs through the GW emission.

Regardless of such different formation scenarios, however, there
should be a much more abundant population of their progenitor BBHs.
Since those BBHs would have wider separations and thus longer orbital
periods, they do not generate a detectable GW signal, except for the
last few seconds before the final merger. Also they are difficult to
be directly imaged unless they are surrounded by appreciable accretion
disks. Therefore, the presence of such unseen binaries have to be
searched for through their dynamical influence on nearby visible
objects, including a radial velocity variation of an outer tertiary
star due to the inner BBH.  Such a detection is challenging, but if
successful, it will not only constrain the formation and evolutionary
channel towards the GW emitting BBHs, but also establish a yet unknown
class of exciting astrophysical objects.

Indeed there exist a couple of known interesting systems that are
relevant for such a strategy. One is a triple system consisting of a
white dwarf-pulsar binary and an outer white dwarf orbiting around the
inner binary \citep{Ransom2014}. The system was detected with the
arrival time analysis of the pulsar. Quite interestingly, the inner
and outer orbits of the triple system are near-circular and coplanar;
the eccentricities of the inner and outer orbits are $e_\ii\sim
6.9\times 10^{-4}$ and $e_\oo\sim 3.5 \times 10^{-2}$, and their
mutual inclination is $i = (1.20\pm 0.17)\times 10^{-2}~\mathrm{deg}$.

The other is a red giant 2M05215658+4359220 with an unseen massive
object of $\sim 3M_\odot$, possibly a BH or even a neutron
star binary \citep{Thompson2019}.  The system was discovered from a
systematic survey of stars exhibiting anomalous accelerations. The
follow-up radial-velocity (RV) observation indicates that the orbit is
also near-circular; $e_\oo=0.00476\pm0.00255$.  These two examples are
very encouraging, implying that the dynamical search for unseen
companions of visible objects is very rewarding.

A very recent discovery of a star--BH binary system, LB-1, by
\citet{Liu2019} is added in the above list. Later, \citet{Abdul-Masih2019} and \citet{El-Badry2019} pointed out that the mass
  of the inner BH would plausibly be $5 M_\odot \lesssim M \lesssim 20 M_\odot$, much smaller than the
  original claim of $\sim 70 M_\odot$ by \citet{Liu2019}.  In
  addition, \citet{Shen2019} put a strong constraint on the presence
  of an inner BBH for LB-1.  Nevertheless, it is very encouraging in a
  sense that such a new population of star-BH binary systems in a
  nearly circular orbit really exists, perhaps abundantly in our
  Galaxy.

Indeed, there are several on-going/future projects that search for
unseen companions around stars.  {\it Gaia} was launched at the end of
2013, and performs astrometric survey for a billion of stars in the
Galaxy. Since the astrometry of {\it Gaia} has great precision
especially for bright stars, it can detect a subtle motion of a star
around an unseen object. Therefore, in addition to the survey of
stellar binaries \citep[e.g.][]{El-Badry2018,Ziegler2018}, there are
many proposals to search for star -- BH binaries with {\it Gaia}
\citep[e.g.][]{Breivik2017,Kawanaka2017,Mashian2017,Yamaguchi2018}.
\citet{Yamaguchi2018}, for instance, estimate that {\it Gaia} can
detect $200-1000$ binaries in its 5 year operation considering the
binary evolution scenario.

Additionally, {\it TESS} launched in 2018 is carrying out a
photometric survey of nearby stars to search for transit
planets. \citet{Masuda2019} conclude that {\it TESS} will potentially
discover $\mathcal{O}(10)$ binaries consisting of a star and an unseen
compact object through identifying relativistic effects in their
photometric light-curves, for instance.

Thus it is quite likely that {\it Gaia}, {\it TESS} and other surveys
will detect many binary systems with an unseen object. Given the LIGO
discovery of very tight BBHs, it is natural to expect that a fraction
of those systems are indeed triple systems that host unseen inner
BBHs. Therefore it is important to see if one can dynamically
distinguish between a single BH and a binary BH in such triple
systems. In addition, the stellar evolution of multiple systems is
still poorly understood. Future detection of a triple with an inner BBH would shed light on dynamical processes in triple evolution \citep[e.g.][]{Toonen2016,Hamers2013,Stephan2016}.

\begin{figure*}
  \begin{center}
    \includegraphics[clip,width=12.0cm]{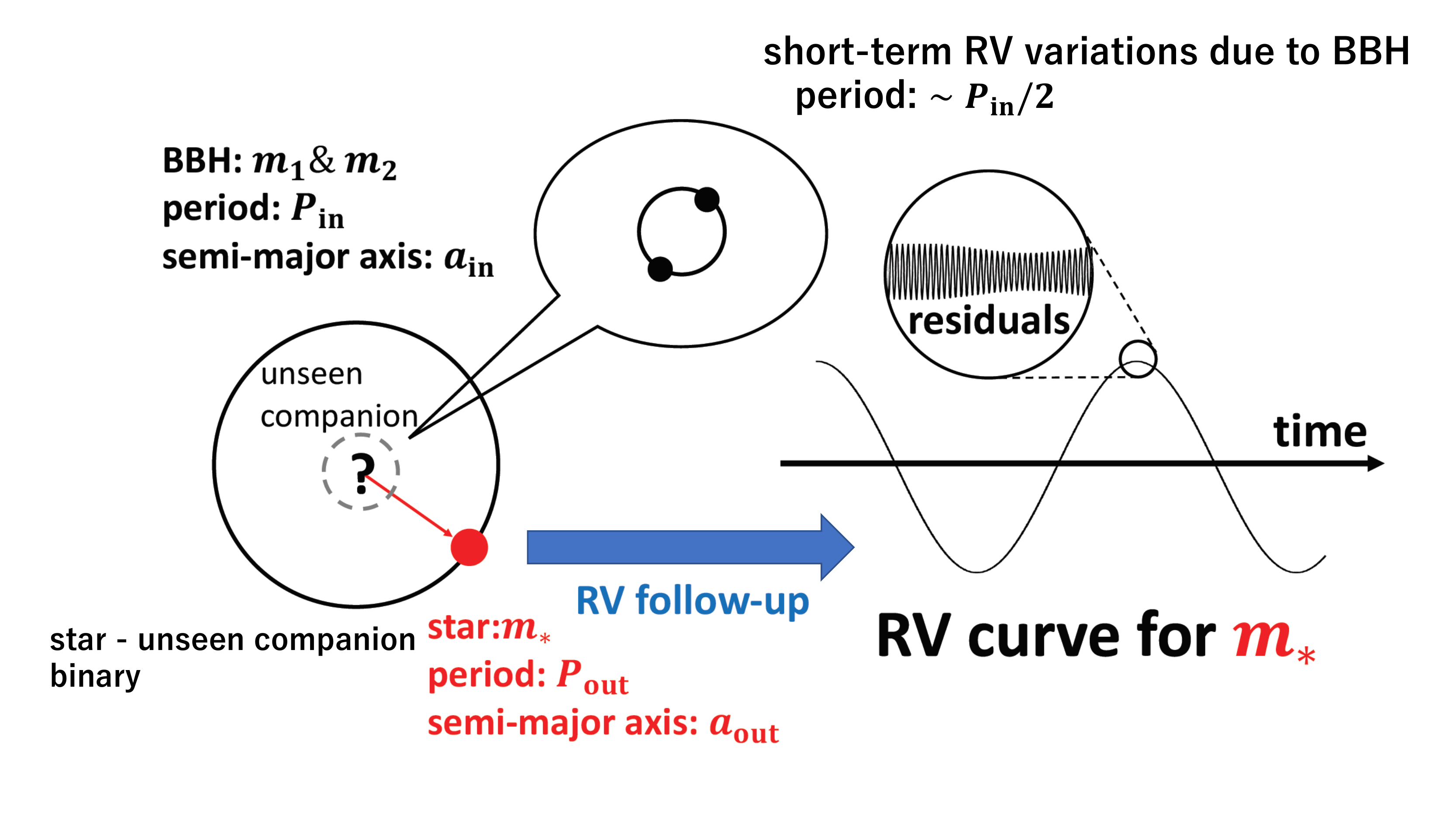}
  \end{center}
\caption{A schematic illustration for a methodology to search for inner
  BBHs using the RV variation of a tertiary star.
\label{fig:illust}}
 \end{figure*}

In this paper, we propose a methodology to search for an inner BBH in
a triple system from the short-term outer stellar RV variations due to
inner BBH perturbation through intensive RV monitoring (see Figure
\ref{fig:illust}). Interestingly, such RV variations were first
considered by \citet{Schneider2006} in a context of a possible
degeneracy between an inner binary and an S-type circumbinary planet.
Later, \citet{Morais2008} derived an analytic perturbative formula for
RV variations in a coplanar circular triple, and pointed out that the
inner binary model produces two nearby but distinct frequency modes
around twice its orbital frequency. If the two modes can be identified
separately, the degeneracy with the planet model, which produces a
single frequency mode, can be broken in principle.  Their work is
further extended for eccentric and inclined triples by
\citep{Morais2011}.

Such approximate analytic formulae are very useful in understanding
the origin and parameter dependence of the RV variations, but they
neglect the back-reaction on the inner BBH by the outer orbit.
Therefore accurate numerical simulations are needed for broader and
more realistic configurations of such triples.  This is what we
conduct in what follows. We also generate mock RV curves, and discuss
the observational feasibility for our fiducial model.

The rest of the paper is organized as follows.  In section
\ref{sec:formula}, we first summarize the analytic approximation RV
formula for a coplanar circular triple by \citet{Morais2008}, and
explain features of the RV variations induced by an inner unseen
binary. As a specific application of the formula, we constrain a
possible inner binary in a star-- unseen-object binary system
2M05215658+4359220. Section \ref{sec:circular} introduces a set of
fiducial triple configurations. In particular, we focus on a coplanar
circular triple, and present a methodology to search for an inner BBH
through the RV curves computed numerically.  Section \ref{sec:mock}
applies the methodology for coplanar eccentric triples, and discusses
the observational feasibility through mock observations. We discuss
how to break the degeneracy of the RV variations due to circumbinary
planets and the inner BBHs.  Finally, section \ref{sec:summary}
summarizes and concludes this paper.  Appendix presents a detailed
derivation of an analytic approximation formula by \citet{Morais2008}
and the parameter degeneracy of the RV variations in the S-type planet
and the inner BBH models.

%%%%%%%%%%%%%%%%%%%%%%%%%%%%%%%%%%%%%%%%%%%%%%%%%%%%%%%%%%%%%%
%%%%%%%%%%%%%%%%%%%%%%%%%%%%%%%%%%%%%%%%%%%%%%%%%%%%%%%%%%%%%%

\section{Radial-velocity variation for a coplanar circular triple
  \label{sec:formula}}

In this section, we first present an approximation formula for RV
variations for a coplanar circular triple with an inner binary, which
was derived by \citet{Morais2008}. This analytic formula is useful to
show basic feature of RV variations, and also to estimate the expected
amplitude of RV variations in designing an RV followup strategy
discussed in the next section. As a specific application for that
purpose, we put a constraint on a possible inner binary in a binary
system 2M05215658+4359220. For definiteness, we summarize a derivation
of the approximate formula in appendix \ref{sec:derivation}.
\subsection{Analytic approximation formula}

We consider a coplanar circular hierarchical triple consisting of an
inner binary and a star.  The RV of the tertiary star orbiting around
the inner binary \citep{Morais2008} is given by
%%%%%%%%%%%%%%%%%%%%%%%%%%%%%%
\begin{eqnarray}
\label{eq:RV}
  V_\mathrm{RV} & \approx & K_0 \left[1+\frac{3}{4}\left(\frac{a_\ii}{a_\oo}\right)^{2} \frac{m_1m_2}{(m_1+m_2)^2}\right] \sin{I}\cos[\nu_\oo t+\theta_0] \nonumber
  \\ & - & \frac{15}{16}K_{\mathrm{BBH}}\sin{I} 
\cos[(2\nu_\ii-3\nu_\oo)t+(2S_0-\theta_0)] \nonumber \\
& + & \frac{3}{16}K_{\mathrm{BBH}}\sin{I}
\cos[(2\nu_\ii-\nu_\oo)t+(2S_0+\theta_0)],
\end{eqnarray}
%%%%%%%%%%%%%%%%%%%%%%%%%%%%%%
where $I$ is the inclination of the stellar orbit axis with respect to
the line of sight, $\nu_\ii$ and $\nu_\oo$ are mean motions of the
inner and outer orbits, and $S_0$ and $\theta_0$ are
constant phases specified by the initial conditions.

The first term in the right-hand-side of equation (\ref{eq:RV}) corresponds to the Keplerian motion of the star.
The semi-amplitude of the unperturbed Keplerian RV for a edge-on view is given by
%%%%%%%%%%%%%%%%%%%%%%%%%%%%%%
\begin{eqnarray}
K_0 \equiv \frac{m_1+m_2}{m_{1}+m_2+m_*} a_\oo \nu_\oo ,
\end{eqnarray}
%%%%%%%%%%%%%%%%%%%%%%%%%%%%%%
where $m_1$, $m_2$ and $m_*$ are the masses of the inner binary and
the outer star, and $a_\oo$ is the semi-major axis of the stellar orbit.
For further detail, see appendix \ref{sec:derivation}.

The second and third terms express the RV variations induced by an
inner binary. The semi-amplitude of the variations is
proportional to $K_\mathrm{BBH}$ defined as
%%%%%%%%%%%%%%%%%%%%%%%%%%%%%%%%%%%%%%
\begin{eqnarray}
\label{eq:KBBH}
K_\mathrm{BBH} \equiv \frac{m_1 m_2}{(m_1+m_2)^2}
\sqrt{\frac{m_{1}+m_2+m_*}{m_{1}+m_2}}
\left(\frac{a_\ii}{a_\oo}\right)^{3.5}K_0
\end{eqnarray}
%%%%%%%%%%%%%%%%%%%%%%%%%%%%%%%%%%%%%%
with $a_\ii$ being the semi-major axis of the inner binary.

Equation (\ref{eq:RV}) indicates that the RV variations due to an
inner binary consist of two periodic terms with frequencies of
$2\nu_\ii -\nu_\oo$ and $2\nu_\ii -3\nu_\oo$.  Since we consider a
system with $\nu_\ii \gg \nu_\oo$, $K_\mathrm{BBH}$ is much smaller
than $K_0$ by a factor of $(a_\ii/a_\oo)^{3.5}$, and their frequencies
are well approximated by $2\nu_\ii$. Thus it is not easy to identify
the two distinct modes observationally.

\citet{Morais2008} pointed out, however, that the separation of the
two modes is crucial to distinguish the RV variation of the inner
binary against a single periodic modulation induced by a planet
orbiting the star.  \citet{Morais2011} extended this consideration for
eccentric and inclined cases using a perturbation theory.

The above result shows that a planet around a star in a binary system
produces a signal very similar to that expected for an inner binary in
a triple system. We discuss the degeneracy between the two models in
section \ref{sec:degeneracy}.  In turn, a detection of an inner binary
requires a well-organized observational monitor with short-cadence as
we discuss below.

\subsection{Constraint on a binary system 2M05215658+4359220
  \label{sec:2M0521}}

\citet{Thompson2019} reported a discovery of a binary system
2M05215658+4359220 that consists of a red giant and an unseen massive
object. They first searched for systems exhibiting anomalously large
radial accelerations from the Apache Point Observatory Galactic
Evolution Experiment (APOGEE) radial velocity data, and selected 200
candidates of such binaries.  After checking the photometric
variations from the All-Sky Automated Survey for Supernovae (ASAS-SN)
data, they identified 2M05215658+4359220 as the most likely binary
candidate.  Subsequently, they performed the radial velocity follow-up
observation with the Tillinghast Reflector Echelle Spectrograph (TRES)
on the $1.5~\mathrm{m}$ telescope at the Fred Lawrence Whipple
Observatory (FLWO).  They obtained 11 spectra with the precision of
about $0.1~\mathrm{km s^{-1}}$ over several months in 2017 and 2018.

Interestingly, the orbital period of the system turned out to be very
close to that of the photometric variation of the star, indicating the
tidal synchronization.  Therefore they assume that the inclination of
the rotation axis of the outer red giant $i_{\mathrm{rot}}$ is equal
to its orbital inclination $i_{\mathrm{orb}}$ with respect to the line
of sight: $i\equiv i_{\mathrm{rot}}=i_\mathrm{ orb}$.  This enabled
them to estimate the best-fit parameters of the system (Table
\ref{tab:par1}) from the RV data and the spectroscopic analysis of the
red giant.

%%%%%%%%%%%%%%%%%%%%%%%%%%%%%%%%%%%%%%%%%%%%%%%%%%%
\begin{table}
\begin{center}
\begin{tabular}{|l|c|c|} \hline
parameter & value & meaning   \\ \hline \hline
$P_\oo$ & $83.205\pm0.064~\mathrm{days}$ & orbital period  \\
$m_\mathrm{co}$ & $3.3^{+2.8}_{-0.7}~\mathrm{M_{\odot}}$
& mass of an unseen companion  \\
$m_{\mathrm{giant}}$ & $3.2^{+1.0}_{-1.0}~\mathrm{M_{\odot}}$
& mass of a red giant   \\
$e_\oo$ & $0.00476\pm 0.00255$ & eccentricity    \\
$\omega_\oo$ & $197.13\pm32.07~\mathrm{deg}$
& argument of pericenter    \\ 
$\sin{i}$ & $0.97^{+0.03}_{-0.12}$ & inclination of the orbital plane \\
$R_\mathrm{giant}$ & $30^{+9}_{-6}~R_\odot$
& radius of a red giant \\
\hline
\end{tabular}
\caption{Best-fit parameters for the binary system
  2M05215658+4359220 \citep[]{Thompson2019} \label{tab:par1}}
\end{center}
\end{table}
%%%%%%%%%%%%%%%%%%%%%%%%%%%%%%%%%%%%%%%%%%%%%%%%%%%

\citet{Thompson2019} estimated the mass of the unseen companion to be
$m_{\mathrm{CO}}=3.3^{+2.8}_{-0.7}~\mathrm{M_{\odot}}$.  Since the
value exceeds a conventionally accepted range of the maximum mass of
the neutron star, it could be a single BH, or even a binary neutron
star/BH.  For simplicity, we assume that the inner binary is
near-circular and coplanar with the outer orbit, and set
$m_\mathrm{giant}=m_*$ and $m_{\mathrm{CO}} = m_1+m_2$; see Table
\ref{tab:par1}.  Then we apply the RV approximation formula so as to
constrain the viable parameters for the possible inner binary.

\begin{figure*}
\begin{center}
 \includegraphics[clip,width=13.0cm]{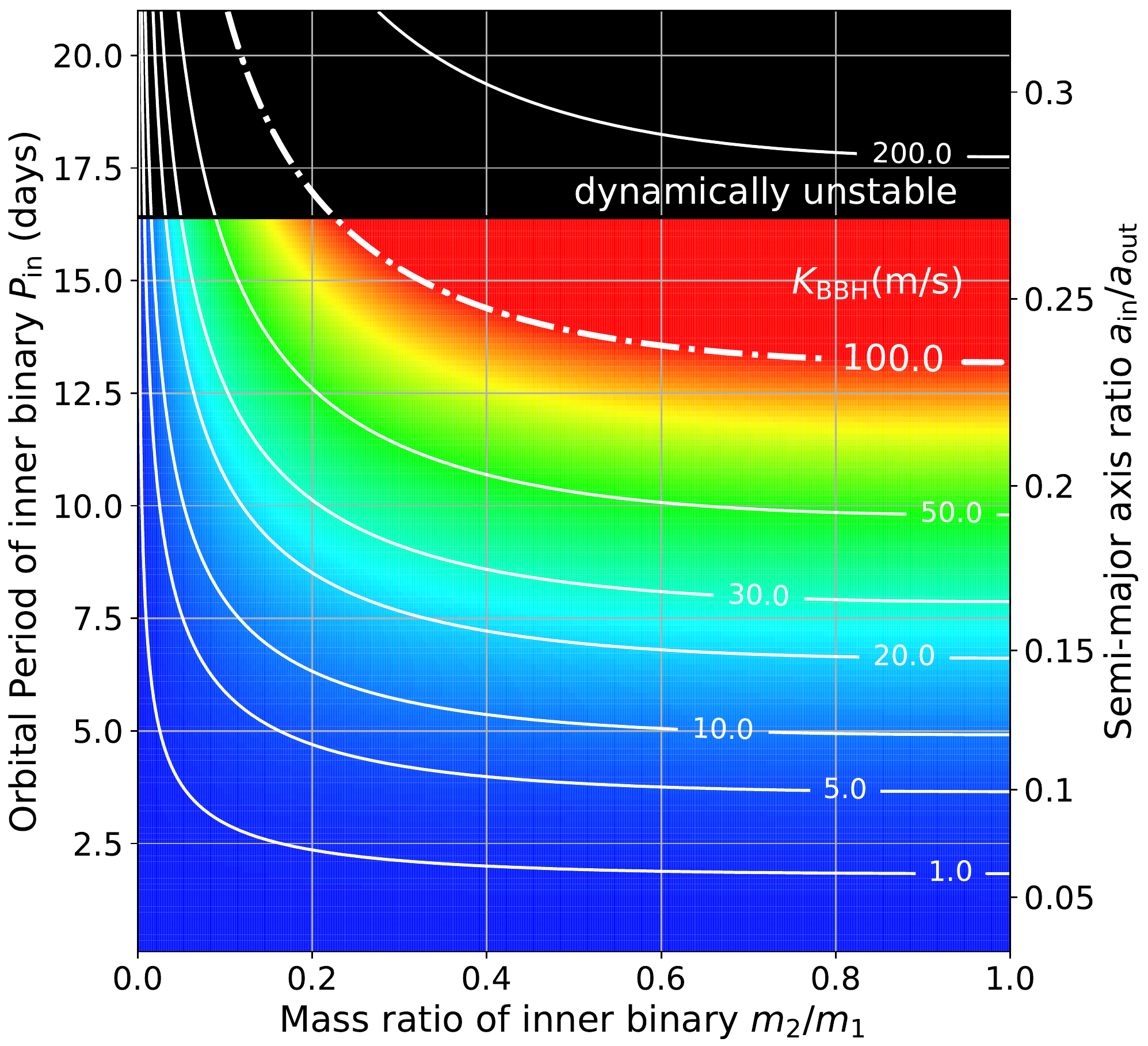}
\end{center}
\caption{Characteristic semi-amplitude of RV variations $K_\BBH$
  induced from a hypothetical inner binary in 2M05215658+4359220. Each
  contour curve is labeled by the value of $K_\BBH$ in units of
  m/s. The black region specifies the dynamically unstable region
  calculated using \citet{Mardling2001}. \label{fig:2M0521}}
\end{figure*}

Figure \ref{fig:2M0521} plots a color contour map of $K_\BBH$ as a
function of the mass ratio $m_2/m_1$ and the orbital period $P_\ii$ of
the possible inner binary for the 2M05215658+4359220 system. We
compute the value of $K_\BBH$ from equation (\ref{eq:KBBH}) using the
parameters listed in Table \ref{tab:par1}.  The color is coded
according to the value of $K_\BBH$ that labels the contour curves.

The axis on the right indicates the semi-major axis ratio
$(a_\ii/a_\oo)$ corresponding to the orbital period $P_\ii$ of the
left axis. Note that the approximation formula is degraded towards the
upper part of Figure \ref{fig:2M0521}. Indeed the three-body system
becomes dynamically unstable if it satisfies \citep{Mardling2001}
%%%%%%%%%%%%%%%%%%%%%%%%%%%%%%%%%%%%%%%%%%%%%%%%%%%%%%%%%%%%%
\begin{equation}
\label{eq:criterion}
\alpha > \left(\frac{a_\ii}{a_\oo}\right)_{\mathrm{crit}}
\equiv \frac{1-e_\oo}{2.8}
\left(\frac{(1+m_3/(m_1+m_2))(1+e_\oo)}{\sqrt{1-e_\oo}}\right)^{-\frac{2}{5}}
\approx 0.270 .
\end{equation}
%%%%%%%%%%%%%%%%%%%%%%%%%%%%%%%%%%%%%%%%%%%%%%%%%%%%%%%%%%%%%
Thus, the perturbation result completely breaks down there as
indicated by the black area in Figure \ref{fig:2M0521}. Although the
approximation formula is degraded for large $(a_\ii/a_\oo)$, we can
still conservatively estimate the semi-amplitude of RV variations in
the allowed region of Figure \ref{fig:2M0521}.

Incidentally, the coalescence time of a circular compact binary
$t_\mathrm{GW}$ via its GW emission is given by \citet{Peters1964}:
%%%%%%%%%%%%%%%%%%%%%%%%%%%%%%%%%%%%%%%%%%%%%%%%%%%%%%%%%%%%%
\begin{equation}
\label{eq:gw}
t_\mathrm{GW} = \frac{5}{256} \frac{c^5 a_\ii^4}{G^3 m_{12}^2 \mu}
\approx 1.88 \times 10^{11} \left(\frac{P_\ii}{\mathrm{day}}\right)^{8/3}
\left(\frac{m_{12}}{M_\odot}\right)^{-5/3}~\mathrm{yrs},
\end{equation} 
%%%%%%%%%%%%%%%%%%%%%%%%%%%%%%%%%%%%%%%%%%%%%%%%%%%%%%%%%%%%%
where $m_{12}=m_1+m_2$, $\mu \equiv m_1m_2/m_{12}$. Equation
(\ref{eq:gw}) ensures that we can safely neglect the effect of GW
emissions unless $P_\ii \ll 1~\mathrm{days}$.

%%%%%%%%%%%%%%%%%%%%%%%%%%%%%%%%%%%%%%%%%%%%%%%%%%%%%%%%%%%%%
%%%%%%%%%%%%%%%%%%%%%%%%%%%%%%%%%%%%%%%%%%%%%%%%%%%%%%%%%%%%%

Considering no detection of anomalous RV variations by TRES beyond
$\sim 100~\mathrm{m/s}$, Figure \ref{fig:2M0521} shows that an assumed
inner coplanar circular binary is constrained to have orbital period
less than a couple of weeks if the binary consists of an almost equal
mass objects.

Assuming a planet orbiting around the red giant with its stellocentric
semi-major axis $a_\mathrm{pl}$, the orbital period of planet
$P_\mathrm{pl}$ is obtained as
%%%%%%%%%%%%%%%%%%%%%%%%%%%%%%%%%%%%%%%%%%%%%%%%%%%%%%%%%%%%%
\begin{eqnarray}
\label{eq:const_pi1}
P_\mathrm{pl} \approx 11~\mathrm{days}
~\left(\frac{a_\mathrm{pl}}{30~\mathrm{R_\odot}}\right)^{3/2}
\gtrsim 11~\mathrm{days}.
\end{eqnarray} 
%%%%%%%%%%%%%%%%%%%%%%%%%%%%%%%%%%%%%%%%%%%%%%%%%%%%%%%%%%%%%
Thus, the period of RV variations by a planet needs to be longer than
$11$ days. On the other hand, the period of variations by an inner
binary is constrained to be shorter than $P_\ii/2\sim 13/2$ days from
Figure \ref{fig:2M0521}.  Since the outer orbiting star is a red giant
of a radius $\sim 30~R_\odot$ (see Table \ref{tab:par1}), we can
basically break the degeneracy between an inner binary and an S-type
circumbinary planet even if the periodic RV variation is detected.

\section{Strategy to search for an inner binary: case for a coplanar circular triple \label{sec:circular}}

In this section, we concentrate on coplanar circular triples before
considering coplanar eccentric systems in the next section.  The main
purpose of this section is to describe our methodology to search for a
possible inner binary in a star -- BH system, and also to check the
accuracy of our numerical simulation against the analytic solution.
Since there are many parameters characterizing a triple system, we
cannot explore an entire parameter space.  Instead, we fix a fiducial
model, and examine the feasibility of our method for the model.

%%%%%%%%%%%%%%%%%%%%%%%%%%%%%%%%%%%%%%%%%%%%%%%%%%%%%%%%%%%
\begin{table}
\begin{center}
\begin{tabular}{|l|c|} \hline
parameter & initial value    \\ \hline \hline
semi-major axis $a_\ii$ & $1.0~\mathrm{au}$ \\
semi-major axis $a_\oo$ & $4.0~\mathrm{au}$ \\
mass of the primary $m_1$ & $10~\mathrm{M_{\odot}}$ \\
mass of the secondary $m_2$ & $10~\mathrm{M_{\odot}}$ \\
mass of the tertiary $m_*$ & $1~\mathrm{M_{\odot}}$ \\
inclinations $I_\ii,I_\oo$ & $90~\mathrm{deg}$ \\ 
arguments of pericenter $\omega_\ii,\omega_\oo$ & $0~\mathrm{deg}$ \\ 
longitudes of ascending node $\Omega_\ii,\Omega_\oo$ & $0~\mathrm{deg}$ \\ 
true anomaly $f_\ii$ & $30~\mathrm{deg}$ \\ 
true anomaly $f_\oo$ & $120~\mathrm{deg}$ \\ 
\hline
\end{tabular}
\caption{Initial values of parameters for our fiducial
  triple. \label{tab:par}}
\end{center}
\end{table}

\begin{table}
\begin{center}
\begin{tabular}{|l|c|} \hline
parameter & value    \\ \hline \hline
orbital period $P_\ii$ & $81.7~\mathrm{days}$ \\
orbital period $P_\oo$ & $638~\mathrm{days}$ \\
semi-amplitude of Keplerian RV $K_0$ & $65.0~\mathrm{km/s}$ \\
variation period (2nd term in eq.(\ref{eq:RV})) $P_\mathrm{var1}$ & $50.5~\mathrm{days}$ \\
variation period (3rd term in eq.(\ref{eq:RV})) $P_\mathrm{var2}$ & $43.6~\mathrm{days}$ \\
half an inner period $P_\ii/2$ & $40.9~\mathrm{days}$ \\
characteristic semi-amplitude of variations $K_\BBH$ & $130.1~\mathrm{m/s}$ \\ 
\hline
\end{tabular}
\caption{Predicted values for orbital motions calculated from equation
  (\ref{eq:RV}) and Table \ref{tab:par}.
\label{tab:predicted}}
\end{center}
\end{table}
%%%%%%%%%%%%%%%%%%%%%%%%%%%%%%%%%%%%%%%%%%%%%%%%%%%%%%%%%%%

Table \ref{tab:par} summarizes the initial values of parameters in our
fiducial model. As before, we use subscripts ``$\ii$" and ``$\oo$" to
denote the parameters for inner and outer orbits, respectively.  We
adopt a $1~M_\odot$ star as the tertiary for simplicity.  Also we
assume a relatively large value for $(a_\ii/a_\oo)$ within the
dynamically stable range \citep{Mardling2001}.  In the present
section, we consider a circular system, and set inner and outer
eccentricities $e_\ii$, $e_\oo$ to $10^{-5}$. We vary this value when we
consider eccentric systems later in section \ref{sec:mock}. As a
reference, Table \ref{tab:predicted} lists the predicted values for
orbital motions calculated from equation (\ref{eq:RV}).

Then, we perform N-body simulations for the coplanar circular triple
with initial conditions in Table \ref{tab:par}, using a public N-body
package {\tt REBOUND} \citep{Rein2012}.  We use WHFast integrator
\citep{Rein2015}, which is one of the fast and accurate symplectic
integrators, and set the calculation time interval to be
$10^{-6}~\mathrm{yr}/2\pi$.  We output the RV data in $0.1$-day cadence
over 800 days in total.

Since we are looking for a tiny RV variation relative to a much larger
Keplerian component, we need to determine the base-line Keplerian
motion accurately and subtract it from the total RV.  Although it may
seem a trivial and straightforward procedure, it is not the case in
reality.

\begin{figure*}
\begin{center}
\includegraphics[clip,width=11.0cm]{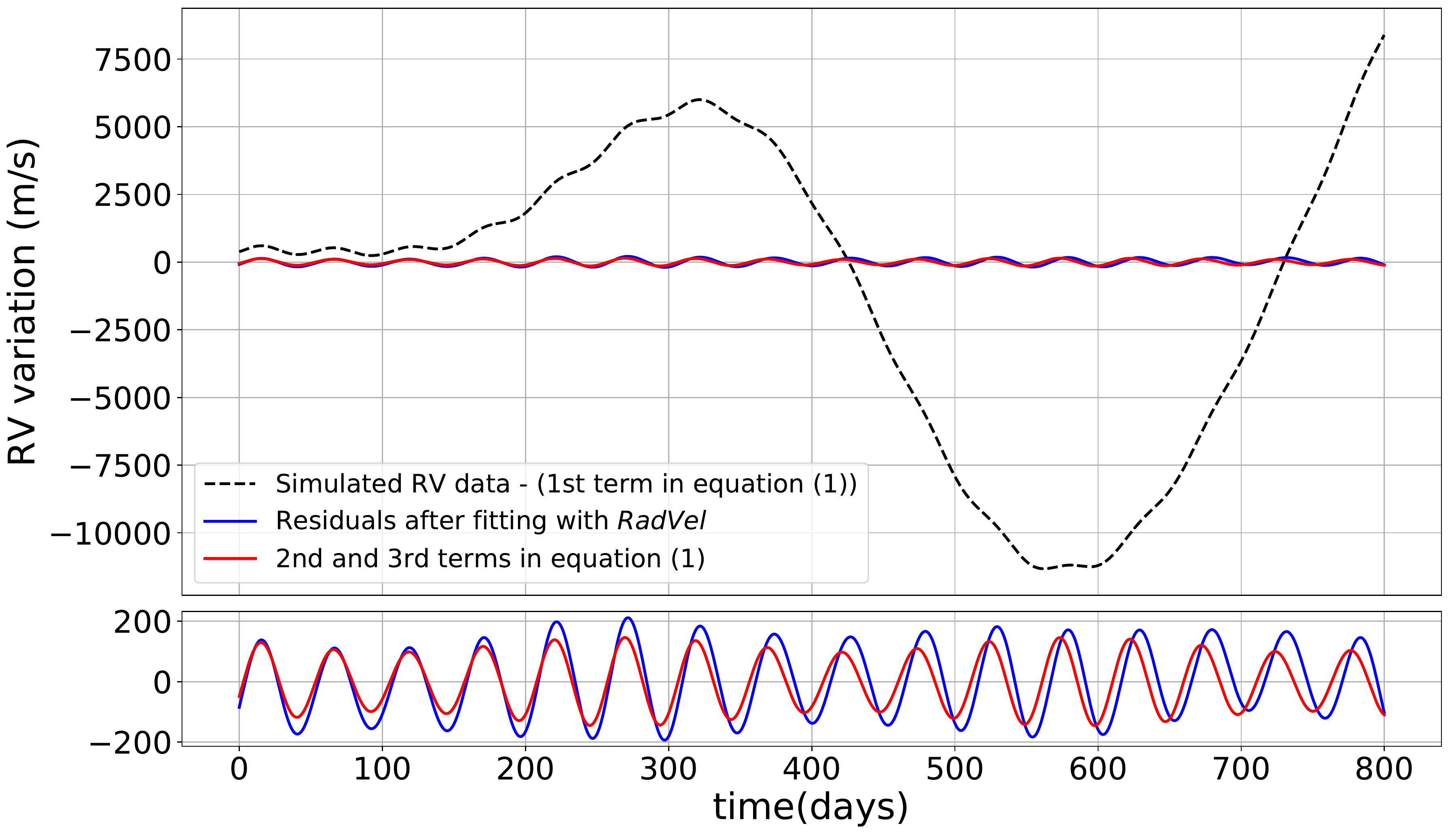}
\end{center}
\caption{Comparison between the analytic approximation 
  (the second and third terms of equation (\ref{eq:RV}) and
  the numerically computed RV residual; see the main text for detail
  \label{fig:comp_c}}
\end{figure*}

Figure \ref{fig:comp_c} illustrates an example of RV residuals.  First
we compute the total RV signal of the tertiary star numerically.
Since we know the initial values of the input parameters (Table
\ref{tab:par}), we subtract the first-term in the right-hand-side of
equation (\ref{eq:RV}) assuming $P_\oo=638$ days and $e_\oo=0$ (Table
\ref{tab:predicted}) from the total RV signal.  The result is plotted
as a black dashed line, which exhibits a large-amplitude modulation of
a period close to $P_\oo$.  This discrepancy comes from the fact that
the orbital period of the outer star is indeed affected by the motion
of the inner binary. In other words, $P_\oo$ is not a constant unlike
as assumed in the approximate formula, equation (\ref{eq:RV}).

Therefore, we next fit the total RV signal using a RV fitting code
{\tt RadVel} \citep{Fulton2018}, and determine the best-fit Keplerian motion
locally over the over 800 days. In doing so, we fix a constant motion
($\gamma$ velocity) and neglect the RV jitters.  The resulting
best-fit values are summarized in Table \ref{tab:fit1}.  The blue line
in Figure \ref{fig:comp_c} is the RV residual after removing the {\it
  locally fitted} Keplerian component with $\tilde{P}_\oo \approx 619.3$ days and $\tilde{e}_\oo\approx 0.016$.  
It is in good agreement with the RV variation induced by an inner binary, the second and third
terms in equation (\ref{eq:RV}), which is plotted as a red line in
Figure \ref{fig:comp_c}.

The above result implies that the removal of the accurate Keplerian
component from the total RV signal is crucial to extract the RV
variations due to an inner binary.  The good agreement between our
numerical result and the analytic formula for the RV variation
simultaneously justifies the robustness of our methodology proposed
here.  There is a tiny discrepancy between the blue and red lines, but
it is likely due to the limitation of the perturbative approximation,
because our fiducial set of parameters are not fully in the
perturbative regime.

Therefore, in what follows, we use fitting with {\tt RadVel} to
determine the base-line Keplerian component, and define the residuals
as RV variations.

%%%%%%%%%%%%%%%%%%%%%%%%%%%%%%%%%%%%%%%%%%%%%%%%%%%%
\begin{table}
\begin{center}
\begin{tabular}{|l|c|} \hline
parameter & value    \\ \hline \hline
orbital period $\tilde{P}_\oo$ & $619.3041~\mathrm{days}$ \\
time of conjunction $\tilde{T}_{\mathrm{j},\oo}$ & $566.12833~\mathrm{days}$ \\
eccentricity $\tilde{e}_\oo$ & $0.01565794$ \\
pericenter argument $\tilde{\omega}_\oo$ & $-1.007588~\mathrm{rad}$ \\
the semi-amplitude of RV $\tilde{K}_0\sin{\tilde{I}}$ & $65931.314~\mathrm{m/s}$ \\ \hline
\end{tabular}
\caption{Best-fit parameters estimated with {\tt RadVel} for our
  fiducial coplanar circular triple. Tildes on each variable indicates
  that they are fitted from the total RV curve over 800 days.
  \label{tab:fit1}}
\end{center}
\end{table}
%%%%%%%%%%%%%%%%%%%%%%%%%%%%%%%%%%%%%%%%%%%%%%%%%%%%

%%%%%%%%%%%%%%%%%%%%%%%%%%%%%%%%%%%%%%%%%%%%%%%%%%%%
%%%%%%%%%%%%%%%%%%%%%%%%%%%%%%%%%%%%%%%%%%%%%%%%%%%%

\section{Mock observation of a coplanar eccentric triple \label{sec:mock}}

This section considers effects of eccentricities for coplanar
triples. Although \citet{Morais2011} provide analytic perturbative RV
formulae even for non-coplanar eccentric cases, the applicability of
the result is limited by the validity of the perturbative
approach. Also the formulas are fairly complicated and it is not easy
to understand the overall behavior of RV variations.  Therefore, we
first provide examples for the predicted RV variations from numerical
simulations, and then perform mock observations including noise and
finite sampling.  Finally we discuss the feasibility of the detection
of an inner binary based on our methodology.

\subsection{RV variations for coplanar eccentric triples
  \label{subsec:anomaly}}

For definiteness, we consider four cases of different eccentricities:
$(e_\ii,e_\oo)=(10^{-5},10^{-5})$, $(10^{-5},0.2)$, $(0.1,10^{-5})$,
and $(0.1,0.2)$. The other parameters are fixed as those listed in
Table \ref{tab:par}.  As in section \ref{sec:circular}, we generate
the RV data in 0.1-day cadence over 800 days using {\tt REBOUND}, and
extract the base-line Keplerian motion fitted with {\tt RadVel}

\begin{figure*}
\begin{center}
\includegraphics[clip,width=17.0cm]{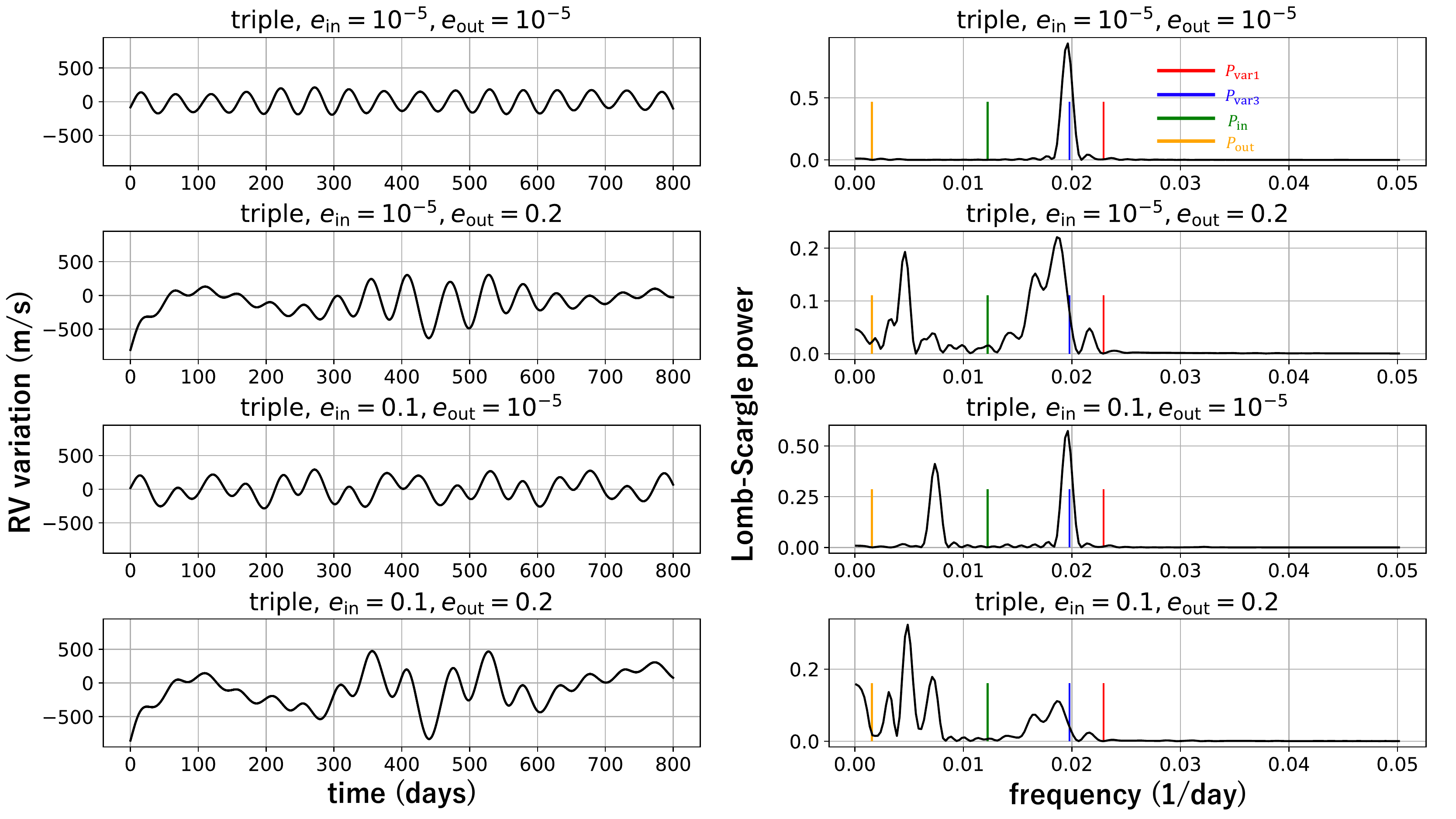}
\end{center}
\caption{RV variations from numerical simulations: RV variation curves
  ({\it left}) and their Lomb-Scargle periodograms ({\it right}). For
  reference, the vertical lines in the right panel indicate relevant
  frequencies of the systems calculated from Table
  \ref{tab:predicted}. \label{fig:templates}}
\end{figure*}

Figure \ref{fig:templates} shows the RV variations ({\it left}) and
the corresponding Lomb-Scargle (LS) periodogram ({\it right}). The
LS periodograms are calculated using a community-developed
core Python package for Astronomy, {\tt Astropy}
\citep[][]{astropy2013,astropy2018}.

The RV variation for a coplanar circular triple exhibits a clear
periodic signal as shown in the top-left panel of Figure
\ref{fig:templates}. According to equation (\ref{eq:RV}), the
variation should have two distinct frequencies $\nu_{\mathrm{var1}}
\equiv 2\nu_\ii - \nu_\oo$, and $\nu_{\mathrm{var3}} \equiv
2\nu_\ii - 3\nu_\oo$, corresponding to
%%%%%%%%%%%%%%%%%%%%%%%%%%%%%%%%%%%%%
\begin{eqnarray}
P_{\mathrm{var1}} &=& \frac{P_\ii P_\oo}{2P_\oo -P_\ii}, 
\end{eqnarray}
%%%%%%%%%%%%%%%%%%%%%%%%%%%%%%%%%%%%%
and
%%%%%%%%%%%%%%%%%%%%%%%%%%%%%%%%%%%%%
\begin{eqnarray}
P_{\mathrm{var3}} &=& \frac{P_\ii P_\oo}{2P_\oo -3P_\ii},
\end{eqnarray}
%%%%%%%%%%%%%%%%%%%%%%%%%%%%%%%%%%%%%
respectively. The LS periodograms in the top-right panel of Figure
\ref{fig:templates} indicates that the RV variation that we compute
numerically is dominated by the mode with $\sim 1/P_{\mathrm{var3}}
\approx 0.02~ {\rm day}^{-1}$, and the mode with $\sim
1/P_{\mathrm{var1}} \approx 0.023~ {\rm day}^{-1}$ is barely visible.
For comparison, we also computed the LS periodogram directly for the
second and third terms of equation (\ref{eq:RV}) with the same
parameters and 0.1-day cadence over 800 days.  We found that the
secondary peak at $\nu_{\mathrm{var1}}$ is not visible either, but
correctly reproduced only when the total base-line duration is
significantly larger than the 800 days we adopted here.  Therefore,
the identification of the two distinct modes due to an inner binary is
very challenging in practice.

The LS periodograms also imply that non-zero eccentricities generate a
variety of additional frequency modes in the LS periodograms, and that
the value of the frequency $\nu_{\mathrm{var3}}$ is fairly insensitive
to $e_\ii$, but decreases as $e_\oo$ increases.  The sensitivity
of the RV variation curve in time domain, instead of the LS
periodogram, on $e_\ii$ and $e_\oo$ may be useful to estimate those
eccentricities of the system.  Thus we conclude that the presence of
the inner binary itself can be inferred robustly. Also we find that
the eccentricity is useful in distinguishing between the inner binary
and the planetary signals as we discuss in section
\ref{sec:degeneracy}.

%%%%%%%%%%%%%%%%%%%%%%%%%%%%%%%%%%%%%%%%%%%%%%%%%%%%%%%
%%%%%%%%%%%%%%%%%%%%%%%%%%%%%%%%%%%%%%%%%%%%%%%%%%%%%%%

\subsection{Mock observations} \label{subsec:mock_obs}

%%%%%%%%%%%%%%%%%%%%%%%%%%%%%%%%%%%%%%%%%%%%%%%%%%%%%%%
\begin{figure*}
\begin{center}
\includegraphics[clip,width=17.0cm]{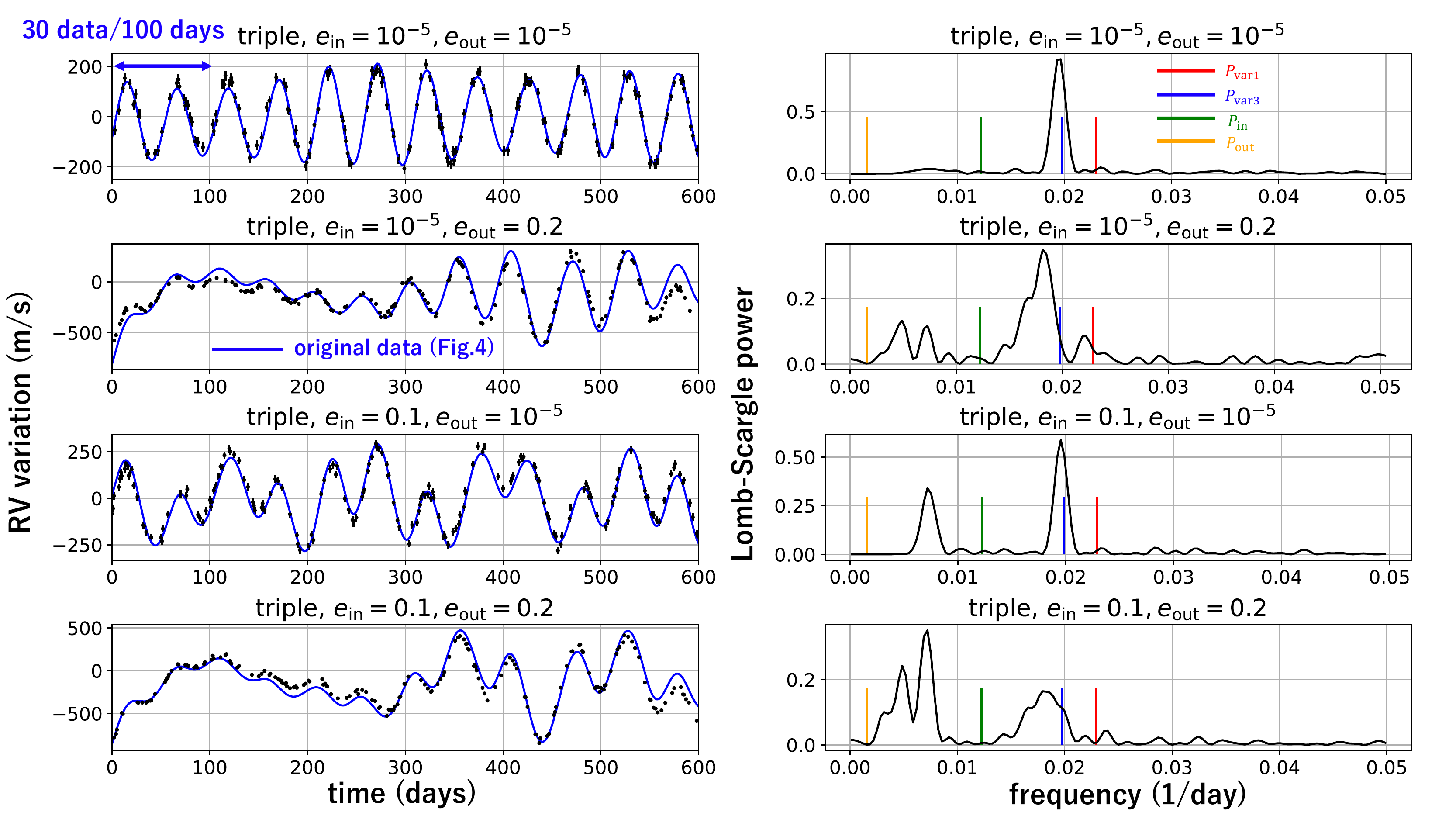}
\end{center}
\caption{RV variation curves from mock observations of the original data
  plotted in Figure \ref{fig:templates} and their LS periodograms.
  We sample 30 data points
  from each 100-days segment of the total RV data,
  add 20 m/s Gaussian noise, and extract the Keplerian component
  fitted to the $30\times 6$ points over 600 days.
  Thin blue curves in the left panels correspond to the original data
  without noise plotted in Figure \ref{fig:templates}.
The vertical lines in the right panels indicate relevant
  frequencies of the systems calculated from Table
  \ref{tab:predicted}.
 \label{fig:RV_residuals_mock}}
%%%%%%	
\begin{center}
\includegraphics[clip,width=17.0cm]{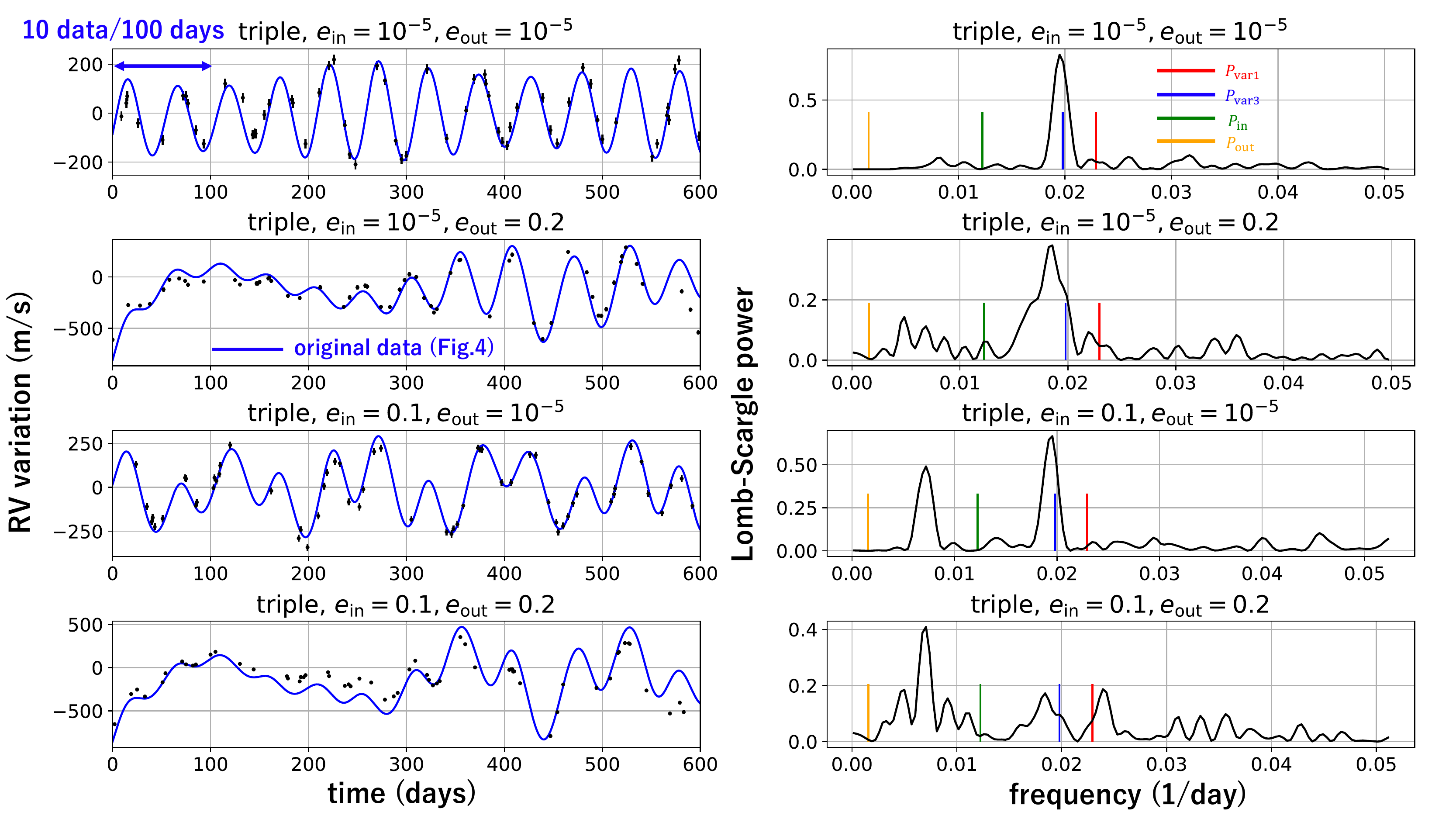}
\end{center}
\caption{Same as Figure \ref{fig:RV_residuals_mock} but sampled 10 data points
  from each 100-days segment. \label{fig:RV_residuals_mock2}}
\end{figure*} 
%%%%%%%%%%%%%%%%%%%%%%%%%%%%%%%%%%%%%%%%%%%%%%%%%%%%%%%

In order to examine the feasibility of our methodology to detect an
inner binary, we perform mock observations of the RV signal including
observational noise and finite sampling effect.  If the tertiary star
is a solar-type star located at $\sim 1~\mathrm{kpc}$, the apparent
magnitude is around $15$. According to \citet{Plavchan2015}, the
statistical RV error is at the level of a few tens of m/s with an
ideal 10m telescope with $\sim 1000$ second exposure. While the noise
is crucially dependent on the nature of the star, we neglect any
systematic/non-Gaussian noise, and add $20$ m/s Gaussian noise into
the RV signal.

The observational cadence also affects the detectability of the inner
binary. We consider two different cases; 10 and 30 data points randomly
selected from one-day cadence data over 100 days.  In both
cases, we assume that the same cadence data are available for 600
days. Then we fit the data over 600 days with {\tt RadVel} to
extract the base-line Keplerian component, and obtain the
RV residuals.

The results are shown in Figure \ref{fig:RV_residuals_mock} for the 30
percent sampling rate, and in Figure \ref{fig:RV_residuals_mock2} for
the 10 percent sampling rate.  Blue lines in the left panels of
Figures \ref{fig:RV_residuals_mock} and \ref{fig:RV_residuals_mock2}
correspond to the original RV variation curves of Figure
\ref{fig:templates}. We note that we add the noise on the {\it total}
RV curve, instead of the RV variation curve (blue lines).  Thus the
extracted Keplerian component is not exactly the same depending on the
Gaussian noise. This is why the data points in the left panels of the
two figures do not necessarily distribute around the blue lines.  This
clearly implies that an accurate determination and extraction of the
base-line Keplerian component is crucial in our methodology.

Except for the limitation, Figures \ref{fig:RV_residuals_mock} and
\ref{fig:RV_residuals_mock2} indicate that the RV variations can
clearly reveal the presence of the inner binary if the relevant
short-cadence RV follow-up is performed.  The LS periodograms in our
mock data are very similar to the idealized cases shown in Figure
\ref{fig:templates}; the peaks around the frequency
$\nu_{\mathrm{var3}}$ are fair robust even in the (modest)
eccentricities we assumed here.

While the inner binary orbital period can be inferred relatively
easily, it would be difficult to determine the mass ratio $m_2/m_1$ of
the binary. As equation (\ref{eq:KBBH}) indicates, the RV variations
are more insensitive to the mass ratio than to the inner binary
separation.  Even a small uncertainty of the amplitude due to noise
would affect the estimate of the mass ratio.

\clearpage

\section{Degeneracy with an S-type circumbinary planet \label{sec:degeneracy}}

%%%%%%%%%%%%%%%%%%%%%%%%%%%%%%%%%%%%%%%%%%%%%%%%%%%%%%%%%%%
\begin{figure}
\begin{center}
\includegraphics[clip,width=5.0cm]{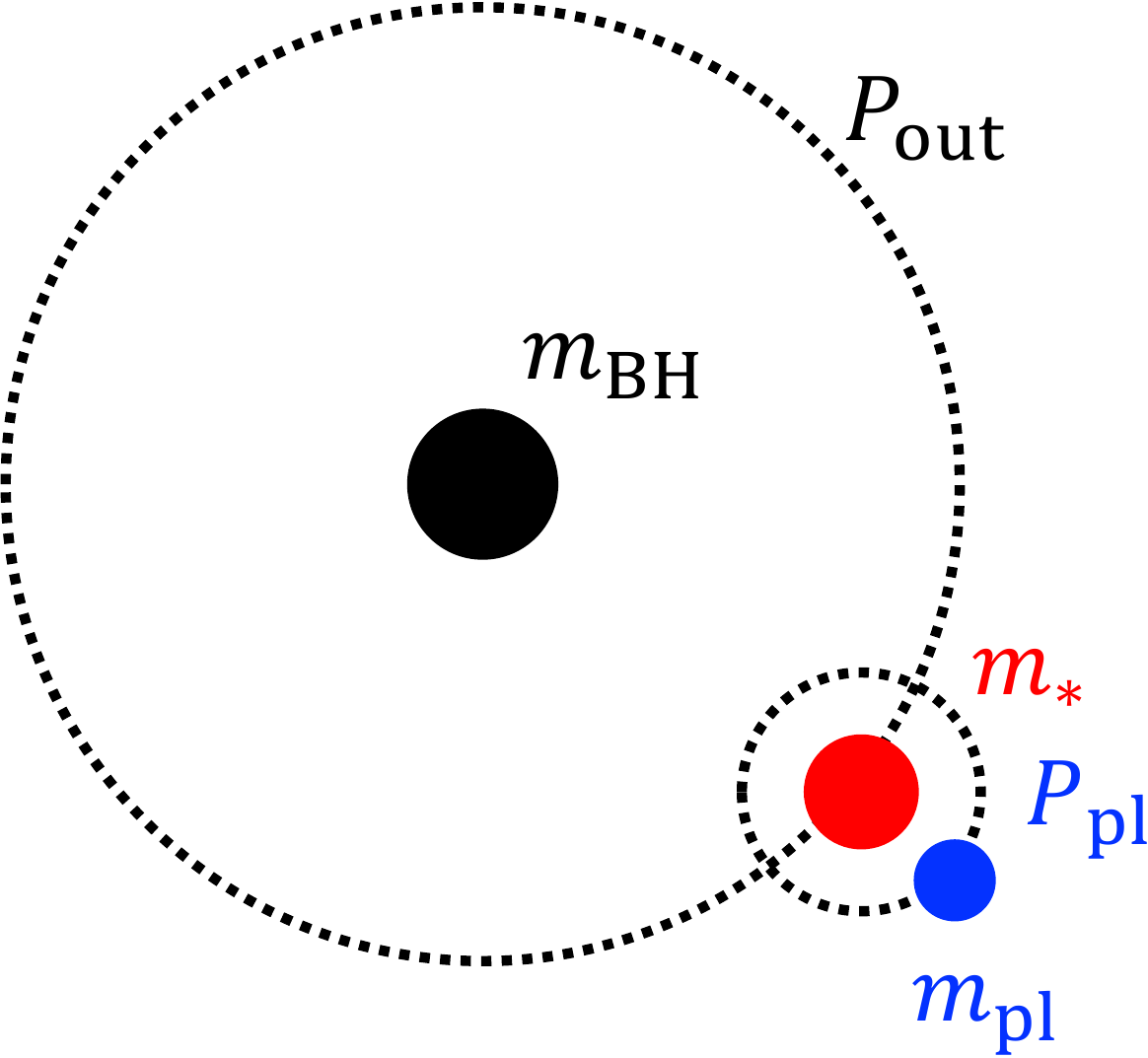}
\caption{A binary consisting of a star and an unseen single BH.
\label{fig:splanet}}
\end{center}
\end{figure}
%%%%%%%%%%%%%%%%%%%%%%%%%%%%%%%%%%%%%%%%%%%%%%%%%%%%%%%%%%%%

%%%%%%%%%%%%%%%%%%%%%%%%%%%%%%%%%%%%%%%%%%%%%%%%%%%%%%%%%%%
\begin{table}
\begin{center}
\begin{tabular}{|l|c|} \hline
parameter & initial value    \\ \hline \hline
semi-major axis $a_\pp$ & $0.27~\mathrm{au}$ \\
semi-major axis $a_\oo$ & $4.0~\mathrm{au}$ \\
mass of the unseen companion $m_\mathrm{BH}$ & $20~\mathrm{M_{\odot}}$ \\
mass of the star $m_*$ & $1~\mathrm{M_{\odot}}$ \\
mass of the planet $m_\pp$ & $2.7~\mathrm{M_{J}}$ \\
inclinations $I_\pp,I_\oo$ & $90~\mathrm{deg}$ \\ 
arguments of pericenter $\omega_\pp,\omega_\oo$ & $0~\mathrm{deg}$ \\ 
longitudes of ascending node $\Omega_\pp,\Omega_\oo$ & $0~\mathrm{deg}$ \\ 
true anomaly $f_\pp$ & $30~\mathrm{deg}$ \\ 
true anomaly $f_\oo$ & $120~\mathrm{deg}$ \\ \hline
\end{tabular}
\caption{Initial values of parameters for a fiducial binary with an S-type
  circumbinary planet orbiting around the star.	\label{tab:parp}}
\end{center}
\end{table}
%%%%%%%%%%%%%%%%%%%%%%%%%%%%%%%%%%%%%%%%%%%%%%%%%%%%%%%%%%%

As mentioned in sections \ref{sec:intro} and \ref{sec:formula}, an
S-type circumbinary planet (see Figure \ref{fig:splanet}) produces a
RV variation similar to that due to an inner binary.  We examine the
degree of a possible degeneracy by performing mock observations
presented in section \ref{sec:mock}.

We consider the configuration of a star-- BH binary system with a
planet around a star. Following Appendix \ref{sec:app_deg}, we choose
a set of parameters for a planet (Table \ref{tab:parp}) so as to mimic
the RV variations produced by our fiducial triple model (Table
\ref{tab:par}).  We also consider non-zero eccentricities of the
planetary and stellar orbits: $(e_\pp,e_\oo)=(10^{-5},10^{-5})$,
$(10^{-5},0.2)$, $(0.1,10^{-5})$, and $(0.1,0.2)$. While we choose the
same set of values for $(e_\ii,e_\oo)$ in subsection
\ref{subsec:mock_obs}, $e_\ii$ and $e_\pp$ do not have to be the same
value in the two different pictures.  We then perform the mock
observations similar in section \ref{sec:mock}. We here consider the
30 percent sampling rate and examine if we can distinguish between the
two pictures observationally.

\begin{figure}
\begin{center}
\includegraphics[clip,width=17.0cm]{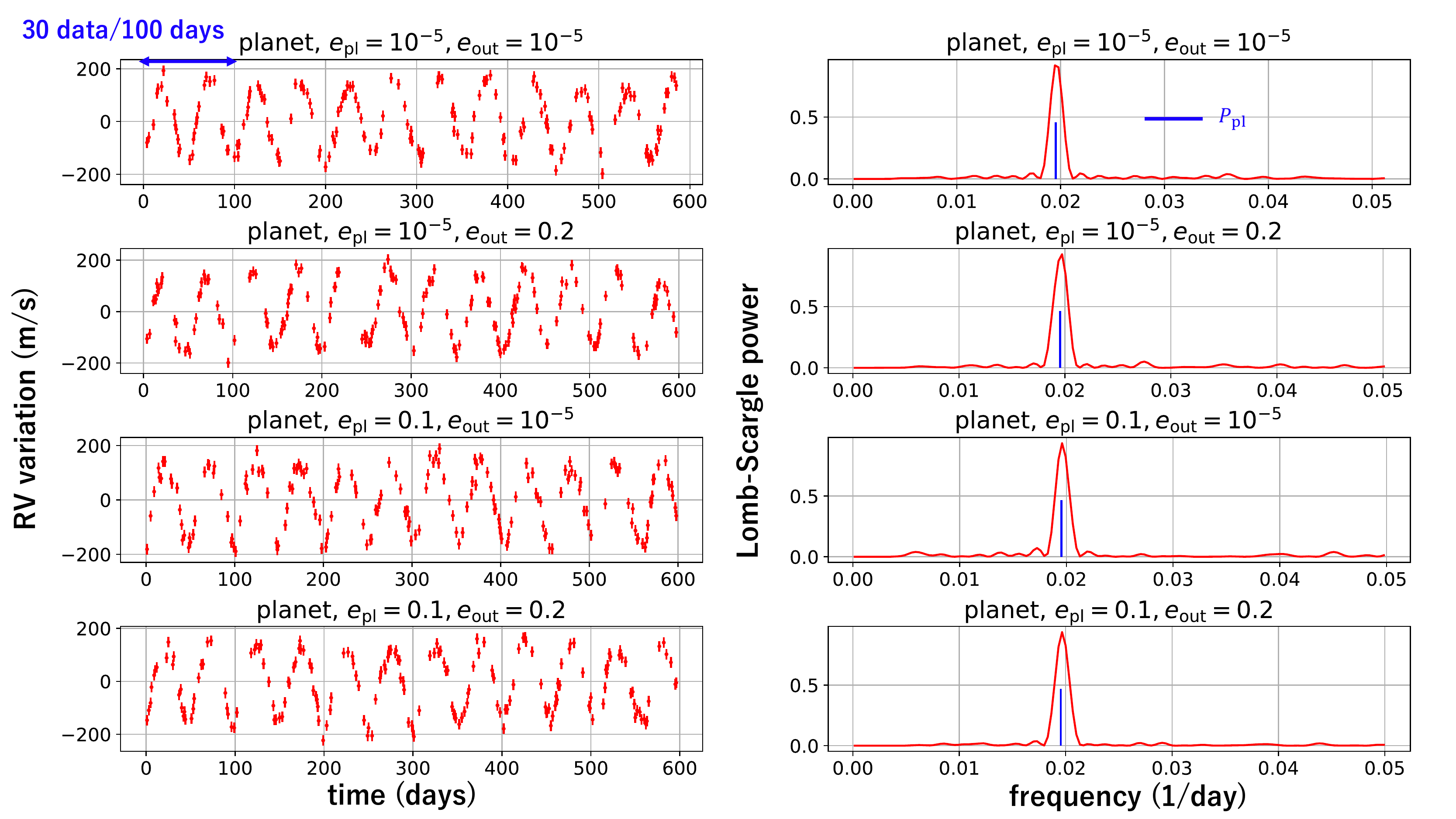}
\end{center}
\caption{RV variations induced by an S-type circumbinary planet
  from mock observation based on the same parameters that are employed in
  Figure \ref{fig:RV_residuals_mock}.
\label{fig:deg}}
\end{figure}

Figure \ref{fig:deg} shows the resulting RV variations ({\it left})
and the LS periodograms ({\it right}) for star--BH binaries with a
planet orbiting around the star.  As expected, a circular inner binary
(top panels in Figure \ref{fig:RV_residuals_mock}) is difficult to be
distinguished from a star with a planet.  The RV variation curves due
to a planet are not so much affected by $e_\pp$ nor $e_\oo$, in
contrast to their sensitivity due to an inner binary on $e_\ii$ and
$e_\oo$ (Figure \ref{fig:RV_residuals_mock}). Thus eccentricities of
the inner binary is indeed helpful to break the degeneracy between the
two models.  In appendix \ref{sec:app_deg}, we provide the parameter
correspondence between the two possibilities for coplanar and circular
systems.

%%%%%%%%%%%%%%%%%%%%%%%%%%%%%%%%%%%%%%%%%%%%%%%%%%%%%%%%%%%
%%%%%%%%%%%%%%%%%%%%%%%%%%%%%%%%%%%%%%%%%%%%%%%%%%%%%%%%%%%

%%%%%%%%%%%%%%%%%%%%%%%%%%%%%%%%%%%%%%%%%%
\section{Summary} \label{sec:summary}
%%%%%%%%%%%%%%%%%%%%%%%%%%%%%%%%%%%%%%%%%%

The LIGO-VIRGO detection of gravitational waves implies that the
universe harbors a number of massive BBHs that have been dismissed in
traditional astronomy.  Since the GW emission from such BBHs is
detectable only for the last few seconds before the final merger, their
progenitors should be even more numerous and remain undetected as a
population of unseen binaries with longer orbital periods.

We propose a methodology to detect such wider-separation BBHs from the
motion of a star orbiting around them. The radial velocity of the
tertiary star in a star -- BBH triple system exhibits a periodic
variation due to the orbital motion of the inner BBH. The variation
signal is small compared with the Keplerian component of the stellar
motion. Nevertheless it is comparable to, or even larger than, a
typical amplitude of the radial velocities for observed exoplanetary
systems. Thus such a velocity variation signal is indeed detectable
if such star-BBH triples are identified.

There are several on-going projects that search for star --
compact-object binaries based on astrometry, transit and
radial-velocity surveys.  They will soon discover a number of star-BH
binary candidates, and a fraction of such binaries may be star-BBH
triples, instead of mere binaries. Indeed a couple of binary systems,
2M05215658+4359220 \citep{Thompson2019} and LB-1 \citep{Liu2019},
provide particularly interesting examples.

We have generated a series of mock radial velocity curves for coplanar
systems, and showed that a short-period variation signal due to the
inner BBH is detectable relatively easily as long as the star is
sufficiently bright to allow for high-resolution spectroscopy, but
that the mass ratio of the inner BBHs is difficult to estimate. The
eccentricity of the tertiary star is useful in breaking the degeneracy
between a planet orbiting the star and an inner BBH.

We find that a short-period variation in the 2M05215658+4359220
binary, if detected at all, should be due to the inner unseen binaries
(neutron stars and/or BHs), since the red giant in the system
cannot accommodate a planet with short orbital periods.

The discovery of LB-1 strongly indicated a presence of abundant star-BH binary systems in our Galaxy.
LB-1 was originally claimed to contain a $68^{+11}_{-13}M_\odot$ unseen companion, which is intriguingly close to the total mass of several BBHs discovered by LIGO-VIRGO.
Although subsequent studies \citep[e.g.][]{Abdul-Masih2019,El-Badry2019} suggested that the mass of the unseen companion is much smaller, the system still remains a promising target of the future RV follow-up. 
The current analysis focused on coplanar triples, but the inner and outer orbits are more likely to be mutually inclined. The extension of our methodology to LB-1 in a more general configuration will be examined and reported elsewhere.

Finally, we would like to
emphasize that the methodology presented in this paper is no more a mere theoretical
idea. It will find a realistic application in the near future that
might unveil a hitherto unknown population of astronomical objects in
a complementary manner to gravitational-wave observations.

\acknowledgments

We would like to thank Alexandre C. M. Correia for letting us know of
their important work on the degeneracy between the inner binary and
the star with a planet, and Zheng Zheng for useful correspondences on
LB-1.  Simulations and analyses in this paper made use of {\tt REBOUND},
{\tt RadVel}, and {\tt Astropy}.  We gratefully acknowledge the support from
Grants-in-Aid for Scientific Research by Japan Society for Promotion
of Science (JSPS) No.18H01247 and No.19H01947, and from JSPS
Core-to-core Program ``International Network of Planetary Sciences''.

\vspace{5mm}

\software{Astropy \citep[][]{astropy2013,astropy2018}, 
          RadVel \citep{Fulton2018},
          REBOUND \citep{Rein2012}
}

\clearpage

\appendix

%%%%%%%%%%%%%%%%%%%%%%%%%%%
\begin{figure*}
\begin{center}
\includegraphics[clip,width=12.0cm]{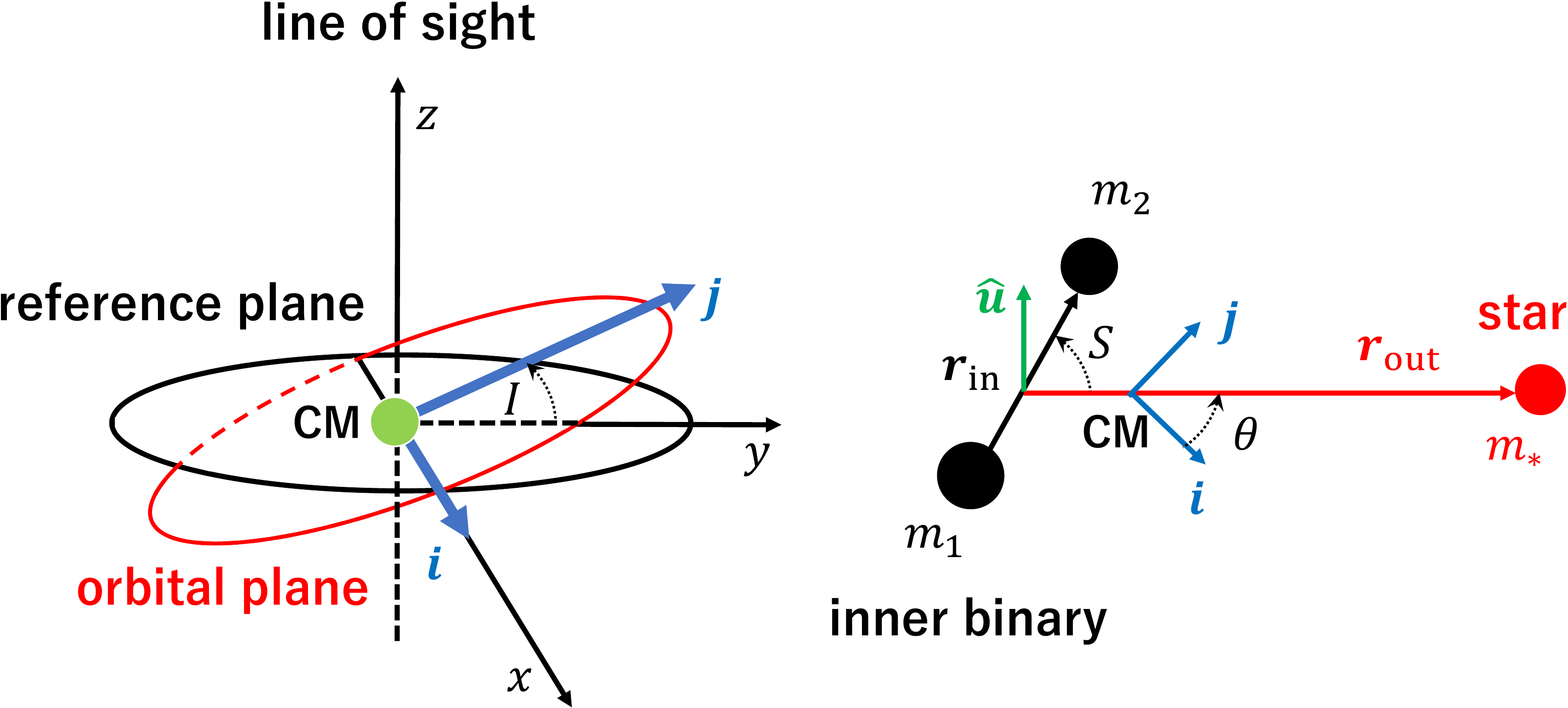}
\end{center}
\caption{A schematic illustration of a triple system consisting of an
  inner binary and a tertiary star following \citet{Morais2008}. The
  left panel shows the common orbital plane of the inner binary and
  the tertiary star. The center of mass of the triple is denoted as
  CM. An observer is located along the $z$-direction with $I$ being
  the inclination of the orbital axis measured by the observer. The
  right panel shows the definitions of orbital parameters of the
  system that are described in the main text.
   \label{fig:schematic}}
\end{figure*}
%%%%%%%%%%%%%%%%%%%%%%%%%%%

\section{The derivation of the RV variations induced by an inner binary}
\label{sec:derivation}

We here derive the equation (\ref{eq:RV}) following
\citet{Morais2008}.  We consider a coplanar circular triple consisting
of an inner binary ($\rho\equiv|\bm{r}_\ii|/|\bm{r}_\oo|\ll1$) and a
star orbiting around the binary (see the right panel of
Fig.\ref{fig:schematic}).  The Hamiltonian of the system to the second
order of $\rho$ is written as
%%%%%%%%%%%%%%%%%%%%%%%%%%%
\begin{eqnarray}
\mathcal{H} = \frac{1}{2}\frac{\bm{p}_\ii^2}{\mu} + \frac{1}{2}\frac{\bm{p}_\oo^2}{\mu_*} - G \frac{m_1m_2}{r_\ii} -G\frac{m_{12}m_*}{r_\oo}-G\frac{m_1m_2m_*}{r_\oo m_{12}}\rho^2 \frac{1}{2}(3\cos^2{S}-1),
\end{eqnarray} 
%%%%%%%%%%%%%%%%%%%%%%%%%%%
where $S\equiv \angle(\bm{r}_\ii\bm{r}_\oo)$, $m_{12}\equiv m_1+m_2$,
$\bm{p}_\ii$ and $\bm{p}_\oo$ are the momenta corresponding to
$\bm{r}_\ii$ and $\bm{r}_\oo$ in terms of Jacobi coordinates,
respectively.  The reduced masses $\mu$ and $\mu_*$ are defined as
%%%%%%%%%%%%%%%%%%%%%%%%%%%
\begin{eqnarray}
\mu \equiv \frac{m_1m_2}{m_{12}} ~~~\mathrm{and}~~~
\mu_* \equiv \frac{m_{12}m_*}{m_{12}+m_*}.
\end{eqnarray}
%%%%%%%%%%%%%%%%%%%%%%%%%%%
Using the Hamiltonian $\mathcal{H}$, the equations of motion are
obtained from the canonical equations:
%%%%%%%%%%%%%%%%%%%%%%%%%%%
\begin{eqnarray}
\dot{\bm{p}}_\ii = -\frac{\partial\mathcal{H}}{\partial \bm{r}_\ii} ~~~ \mathrm{and} ~~~ \dot{\bm{p}}_\oo = -\frac{\partial\mathcal{H}}{\partial \bm{r}_\oo}.
\end{eqnarray}
%%%%%%%%%%%%%%%%%%%%%%%%%%%
Therefore, the explicit forms of equations of motion for the inner
binary and outer star are
%%%%%%%%%%%%%%%%%%%%%%%%%%%
\begin{eqnarray}
\ddot{\bm{r}}_\ii = -G\frac{m_{12}}{r^3_\ii}\bm{r}_\ii + G\frac{m_*}{r^3_\oo}(3\rho\cos{S}\bm{r}_\oo-\bm{r}_\ii)
\label{eq:eom_rin}
\end{eqnarray}
%%%%%%%%%%%%%%%%%%%%%%%%%%%
and
%%%%%%%%%%%%%%%%%%%%%%%%%%%
\begin{eqnarray}
\ddot{\bm{r}}_\oo = -G\frac{m_{12}+m_*}{r^3_\oo}
\left[ \left(1+\frac{\mu}{m_{12}}\frac{\rho^2}{2}(-3+15\cos^2{S})\right)\bm{r}_\oo-\frac{\mu}{m_{12}}3\rho \cos{S}\bm{r}_\ii
\right],
\label{eq:eom_rout}
\end{eqnarray}
%%%%%%%%%%%%%%%%%%%%%%%%%%%
respectively.

The 0-th order solutions of equations (\ref{eq:eom_rin}) and
(\ref{eq:eom_rout}) for a circular system are as follows:
%%%%%%%%%%%%%%%%%%%%%%%%%%%
\begin{eqnarray}
\bm{r}_\ii^{(0)} = a_\ii \cos{S} \hat{\bm{r}}_\oo + a_\ii \sin{S} \hat{\bm{u}} ~~~ \mathrm{and}~~~
\bm{r}_\oo^{(0)} = a_\oo \hat{\bm{r}}_\oo,
\label{eq:rin}
\end{eqnarray}
%%%%%%%%%%%%%%%%%%%%%%%%%%%
where $S = S_0 + (\nu_\ii - \nu_\oo)t$, and $a_\ii$ and $a_\oo$ are the
semi-major axes of the inner and outer orbits, respectively, with $S_0$
being a constant phase determined by the initial positions. The mean motions $\nu_\ii$ and $\nu_\oo$ are defined as
%%%%%%%%%%%%%%%%%%%%%%%%%%%
\begin{eqnarray}
\nu_\ii \equiv \sqrt{\frac{Gm_{12}}{a^3_\ii}}~~~\mathrm{and}~~~ \nu_\oo \equiv \sqrt{\frac{G(m_{12}+m_*)}{a^3_\oo}}.
\end{eqnarray}
%%%%%%%%%%%%%%%%%%%%%%%%%%%

In terms of the Cartesian coordinate on the orbital plane, unit
vectors $\hat{\bm{r}}_\oo$ and $\hat{\bm{u}}$ are
%%%%%%%%%%%%%%%%%%%%%%%%%%%
\begin{eqnarray}
\hat{\bm{r}}_\oo = \cos{\theta} \bm{i} + \sin{\theta}\bm{j}
~~~\mathrm{and}~~~\hat{\bm{u}} = -\sin{\theta} \bm{i} + \cos{\theta} \bm{j},
\label{eq:base_cart}
\end{eqnarray}
%%%%%%%%%%%%%%%%%%%%%%%%%%%
where $\theta \equiv \theta_0 + \nu_\oo t$, and $\bm{i}$ and $\bm{j}$ are
constant base vectors (see Figure \ref{fig:schematic}), with $\theta_0$ being a constant phase determined by the initial
positions. Substituting equations (\ref{eq:rin}) -
(\ref{eq:base_cart}) into the equation (\ref{eq:eom_rout}), we obtain
%%%%%%%%%%%%%%%%%%%%%%%%%%%
\begin{eqnarray}
\ddot{X} = -\frac{Gm_{12}}{a^2_\oo}
\left[ \left(1+\frac{3}{4}\alpha^2\frac{\mu}{m_{12}}\right)\cos{\theta}+\frac{9}{4}\alpha^2\frac{\mu}{m_{12}}\cos{2S}\cos{\theta} + \frac{3}{2}\alpha^2\frac{\mu}{m_{12}}\sin{2S}\sin{\theta}
\right]    
\end{eqnarray}
%%%%%%%%%%%%%%%%%%%%%%%%%%%
and
%%%%%%%%%%%%%%%%%%%%%%%%%%%
\begin{eqnarray}
\ddot{Y} = -\frac{Gm_{12}}{a^2_\oo}
\left[ \left(1+\frac{3}{4}\alpha^2\frac{\mu}{m_{12}}\right)\sin{\theta}+\frac{9}{4}\alpha^2\frac{\mu}{m_{12}}\cos{2S}\sin{\theta} - \frac{3}{2}\alpha^2\frac{\mu}{m_{12}}\sin{2S}\cos{\theta}
\right],
\end{eqnarray}
%%%%%%%%%%%%%%%%%%%%%%%%%%%
where $X$ and $Y$ are defined as $\bm{r}_\oo = X\bm{i}+Y\bm{j}$.

The solutions of these equations are written as
%%%%%%%%%%%%%%%%%%%%%%%%%%%
\begin{eqnarray}
\left(
\begin{array}{c}
X \\
Y \\
\end{array}
\right)
=
\left(
\begin{array}{c}
a_* \cos{\theta} \\
a_* \sin{\theta} \\
\end{array}
\right)
+
\left(
\begin{array}{cc}
\sin{\theta} & \cos{\theta} \\
-\cos{\theta} & \sin{\theta} \\
\end{array}
\right)
\left(
\begin{array}{c}
\delta_X \sin{2S} \\
\delta_Y \cos{2S} \\
\end{array}
\right),
\end{eqnarray} 
%%%%%%%%%%%%%%%%%%%%%%%%%%%
where
%%%%%%%%%%%%%%%%%%%%%%%%%%%
\begin{eqnarray}
a_* \equiv \frac{m_{12}}{m_{12}+m_*} \left(1+\frac{3}{4}\alpha^2 \frac{\mu}{m_{12}}\right)a_\oo,~~~
\delta_X \approx \frac{3}{8}\frac{\mu}{m_{12}}\alpha^4 a_\oo ~~~\mathrm{and}~~~
\delta_Y \approx \frac{9}{16}\frac{\mu}{m_{12}}\alpha^4 a_\oo.
\end{eqnarray}
%%%%%%%%%%%%%%%%%%%%%%%%%%%
The above expressions for $\delta_X$ and $\delta_Y$ are derived on the assumption of $\nu_\oo/\nu_\ii \ll 1$.

The radial velocity $V_\mathrm{RV}$ is defined as
%%%%%%%%%%%%%%%%%%%%%%%%%%%
\begin{eqnarray}
V_\mathrm{RV} \equiv \dot{z} = \dot{Y}\sin{I}.
\end{eqnarray}
%%%%%%%%%%%%%%%%%%%%%%%%%%%
Therefore, we can write down the following RV approximation formula:
%%%%%%%%%%%%%%%%%%%%%%%%%%%
\begin{eqnarray}
V_\mathrm{RV} & \approx & K_0\left(1+\frac{3}{4}\alpha^2 \frac{\mu}{m_{12}}\right) \sin{I}\cos[\nu_\oo t+\theta_0]\nonumber \\ 
& - & \frac{15}{16}K_{\mathrm{BBH}}\sin{I} 
\cos[(2\nu_\ii-3\nu_\oo)t+(2S_0-\theta_0)] \nonumber \\
& + & \frac{3}{16}K_{\mathrm{BBH}}\sin{I}
\cos[(2\nu_\ii-\nu_\oo)t+(2S_0+\theta_0)],
\end{eqnarray}
%%%%%%%%%%%%%%%%%%%%%%%%%%%
where
%%%%%%%%%%%%%%%%%%%%%%%%%%%
\begin{eqnarray}
K_0 \equiv \frac{m_{12}}{m_{12}+m_*} a_\oo \nu_\oo ~~~\mathrm{and}~~~ K_\mathrm{BBH} \equiv \frac{m_1 m_2}{(m_1+m_2)^2}\sqrt{\frac{m_{12}+m_*}{m_{12}}}\alpha^{3.5}K_0.
\end{eqnarray}
%%%%%%%%%%%%%%%%%%%%%%%%%%%

\section{Correspondence of parameters between a triple and
  S-type circumbinary planet \label{sec:app_deg}}

In section \ref{sec:degeneracy}, we find that for a coplanar circular
triple the RV variations are almost degenerate with those produced by
an S-type circumbinary planet. If a star is a giant, we can sometimes
rule out the degeneracy using the same way as for the system
2M05215658+4359220. However, this is not always applicable.  Thus, we
here consider a parameter correspondence for a coplanar circular case
when we cannot distinguish these two possibilities.

The RV variations are basically characterized by its semi-amplitude
$K_{\rm var}$ and period $P_{\rm var}$.  The latter is equal to
$P_{\rm pl}$ and $P_\ii/2$ in the planet-star and inner BBH
interpretations, respectively.

The semi-amplitude of RV variations induced by an inner binary is
estimated from equation (\ref{eq:KBBH}):
%%%%%%%%%%%%%%%%%%%%%%%%%%%%%%%%%%%%%%%%%%%%%%%%%%%%%%%%%%%%%%
\begin{equation}
K_\mathrm{BBH} = (2\pi G)^{1/3}
\left(\frac{m_{123}}{m_{12}}\right)^{5/3} P_\oo^{-1/3} m_{123}^{1/3}
\gamma^{-2} \left(\frac{P_\ii}{P_\oo}\right)^{7/3},
\label{eq:KBBH_ap}
\end{equation}
%%%%%%%%%%%%%%%%%%%%%%%%%%%%%%%%%%%%%%%%%%%%%%%%%%%%%%%%%%%%%
where $P_\oo$ is the orbital period of the outer star,
$m_{12}=m_1+m_2$, $m_{123}=m_{12}+m_*$, and $\gamma \equiv
\left(\sqrt{m_2/m_1}+\sqrt{m_1/m_2}\right)$.

On the other hand, the semi-amplitude of variations induced by a
coplanar circular planet around the outer star is
%%%%%%%%%%%%%%%%%%%%%%%%%%%%%%%%%%%%%%%%%%%%%%%%%%%%%%%%%%%%%
\begin{equation}
K_\mathrm{pl} = (2\pi G)^{1/3} P_{\rm pl}^{-1/3} m_\mathrm{pl} m_*^{-2/3},
\label{eq:Kpl_ap}
\end{equation}
%%%%%%%%%%%%%%%%%%%%%%%%%%%%%%%%%%%%%%%%%%%%%%%%%%%%%%%%%%%%%
where $m_\mathrm{pl}$ is the mass of planet, and we assume that
$m_\mathrm{pl} \ll m_*$.

We cannot distinguish these two interpretations from observation if
both of them induce the RV variations with similar amplitude and
period. For this case, using equations (\ref{eq:KBBH_ap}) and
(\ref{eq:Kpl_ap}), we can obtain the parameter correspondence
considering $K_\mathrm{var}=K_\BBH = K_\mathrm{pl}$, and
$P_\mathrm{var}=P_\mathrm{pl} = P_\ii/2$:
%%%%%%%%%%%%%%%%%%%%%%%%%%%%%%%%%%%%%%%%%%%%%%%%%%%%%%%%%%%%%
\begin{equation}
\label{eq:mp}
m_\mathrm{pl} = 2^{7/3} m_*^{2/3}
\left(\frac{m_{\mathrm{12}}}{m_\mathrm{12}+m_*}\right)^{5/3}
\gamma^{-2} (m_\mathrm{12}+m_*)^{1/3} \left(\frac{P_\mathrm{var}}{P_\oo}\right)^{8/3}.
\end{equation}
%%%%%%%%%%%%%%%%%%%%%%%%%%%%%%%%%%%%%%%%%%%%%%%%%%%%%%%%%%%%%%
Equation (\ref{eq:mp}) relates the parameters in
the two different interpretations.

\begin{figure}
\begin{center}
\includegraphics[clip,width=12.0cm]{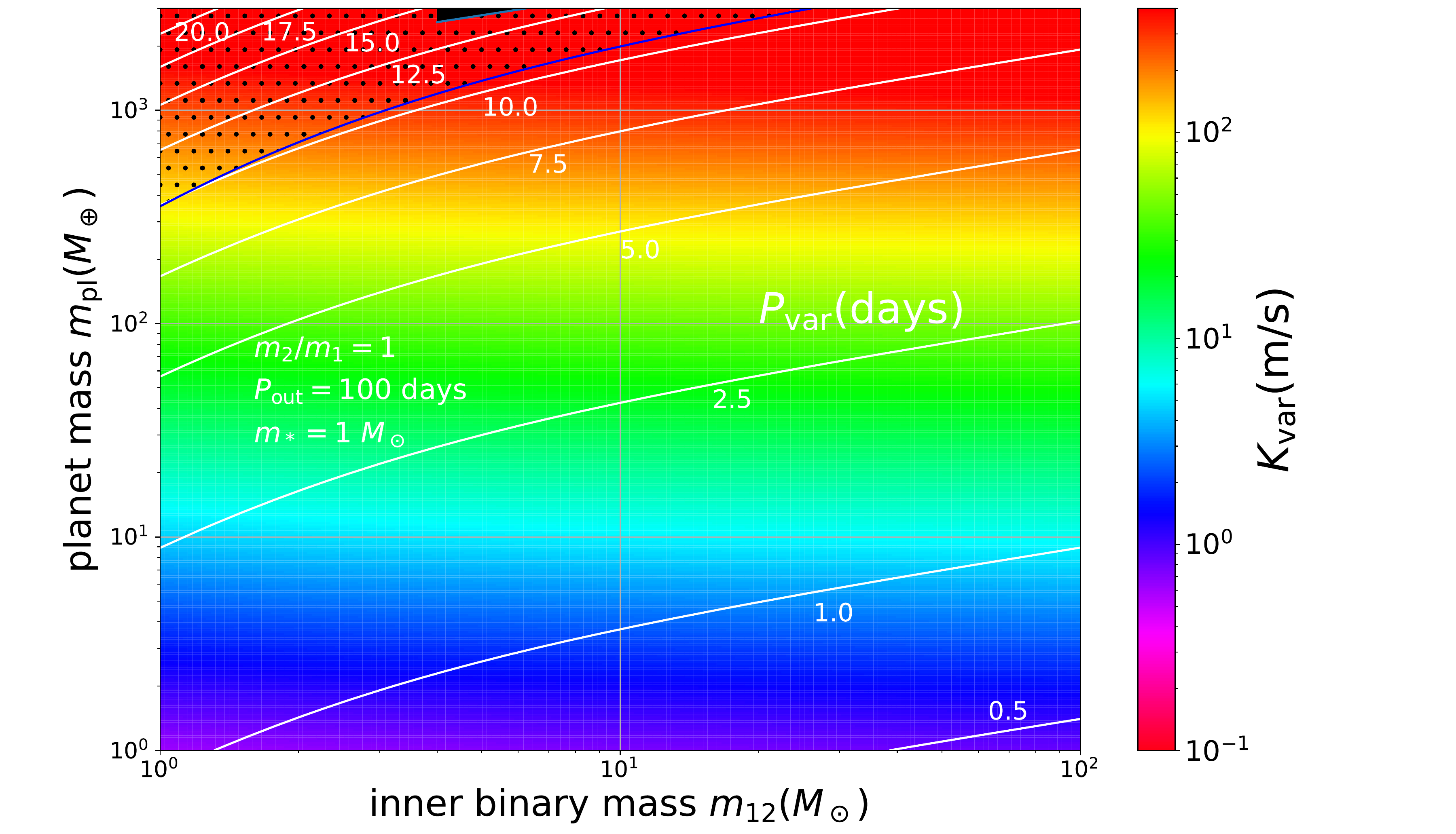}
\caption{An example of the parameter degeneracy between the inner
  binary mass $m_{12}$ and the planet mass $m_{\rm pl}$ that produce
  almost the same RV variation of a period $P_{\rm var}$ and a
  semi-amplitude $K_{|rm var}$. We assume that a tertiary star of
  $1M_\odot$ orbits around an equal-mass BBH with an orbital period of
  $P_\oo = 100$ days.  The dotted region excludes the inner BBH
  picture from the dynamical instability condition for the triple
  system according to \citet{Mardling2001}.  The tiny black region
  excludes the planet picture from inequality
  (\ref{eq:hill}).  \label{fig:const1}}
\end{center}
\end{figure}

Figures \ref{fig:const1} shows the parameter
correspondence between $m_\mathrm{pl}$ in the planet-star
interpretation and $m_{12}$ in the inner BBH interpretation for $P_\oo
= 100~\mathrm{days}$. We assume that $m_2/m_1=1$ and $m_*=1~M_\odot$.  Given the values
of the RV variation semi-amplitude and period, $K_{\rm var}$
(color-coded) and $P_{\rm var}$ (black contour), the corresponding
values of $m_\mathrm{pl}$ and $m_\mathrm{12}$ can be read off from the
figures.

The dotted area above the blue curve is excluded in the inner BBH
picture from the instability of the triple. We use the dynamical
stability criterion \citep{Mardling2001} (see equation
(\ref{eq:criterion})) to compute the region.  The tiny black region is
excluded by the instability of the planet orbiting the star. To compute
the region, we adopt the following Hill instability criterion
\citep[e.g.][]{Holman1999,Barnes2002}:
%%%%%%%%%%%%%%%%%%%%%%%%%%%%%%%%%%%%%%%%%%%%%%%%%%%%%%%%%%%%%
\begin{equation}
\label{eq:hill}
  a_\mathrm{pl} > f \left(\frac{m_*}{3m_\mathrm{12}}\right)^{1/3} a_\oo
  ~~~\left(\mu\equiv\frac{m_{12}}{m_{12}+m_*}\gtrsim 0.8 \right),
\end{equation}
%%%%%%%%%%%%%%%%%%%%%%%%%%%%%%%%%%%%%%%%%%%%%%%%%%%%%%%%%%%%%
where $f=0.36$ is derived from numerical simulations, and
$a_\mathrm{pl}$ is the stellocentric semi-major axis of planet.  Note
that $m_{12}$ is interpreted as the mass of a single BH in a
planet-star picture. Figure \ref{fig:const1} shows that most of regions accept both
interpretations, indicating that a Hot Jupiter around a star in a
star-BH binary system may mimic the RV variation induced by an inner
binary in the system.

%%%%%%%%%%%%%%%%%%%%%%%%%%%
%%%%%%%%%%%%%%%%%%%%%%%%%%%

\end{document}